\documentclass[prl,superscriptaddress,twocolumn]{revtex4}%
\usepackage[ansinew]{inputenc}
\usepackage{graphicx}
\usepackage{amsmath}
\usepackage{amsthm}
\usepackage{bm}
\usepackage{layout}
\usepackage{float}
\usepackage{txfonts}
\usepackage{amsfonts}
\usepackage{amssymb}
\usepackage[squaren]{SIunits}
\usepackage{SIunits}%
\begin{document}

\title{Einstein--Podolsky--Rosen correlations from colliding Bose--Einstein condensates}

\begin{abstract}
We propose an experiment which can demonstrate quantum correlations in a
physical scenario as discussed in the seminal work of Einstein, Podolsky and
Rosen. Momentum-entangled massive particles are produced via the four-wave
mixing process of two colliding Bose--Einstein condensates. The particles'
quantum correlations can be shown in a double double-slit experiment or via
ghost interference.

\end{abstract}
\date{\today}

%

\author{Johannes Kofler}%
%

\affiliation
{Institute for Quantum Optics and Quantum Information (IQOQI), Austrian Academy of Sciences, Boltzmanngasse 3, 1090 Vienna, Austria}%
%

\affiliation
{Max Planck Institute of Quantum Optics (MPQ), Hans-Kopfermannstraße 1, 85748 Garching/Munich, Germany}%
%

\author{Mandip Singh}%
%

\affiliation
{Institute for Quantum Optics and Quantum Information (IQOQI), Austrian Academy of Sciences, Boltzmanngasse 3, 1090 Vienna, Austria}%
%

\affiliation
{Faculty of Physics, University of Vienna, Boltzmanngasse 5, 1090 Vienna, Austria}%
%

\author{Maximilian Ebner}%
%

\affiliation
{Institute for Quantum Optics and Quantum Information (IQOQI), Austrian Academy of Sciences, Boltzmanngasse 3, 1090 Vienna, Austria}%
%

\affiliation
{Faculty of Physics, University of Vienna, Boltzmanngasse 5, 1090 Vienna, Austria}%
%

\author{Michael Keller}%
%

\affiliation
{Institute for Quantum Optics and Quantum Information (IQOQI), Austrian Academy of Sciences, Boltzmanngasse 3, 1090 Vienna, Austria}%
%

\affiliation
{Faculty of Physics, University of Vienna, Boltzmanngasse 5, 1090 Vienna, Austria}%
%

\author{Mateusz Kotyrba}%
%

\affiliation
{Institute for Quantum Optics and Quantum Information (IQOQI), Austrian Academy of Sciences, Boltzmanngasse 3, 1090 Vienna, Austria}%
%

\affiliation
{Faculty of Physics, University of Vienna, Boltzmanngasse 5, 1090 Vienna, Austria}%
%

\author{Anton Zeilinger}%
%

\affiliation
{Institute for Quantum Optics and Quantum Information (IQOQI), Austrian Academy of Sciences, Boltzmanngasse 3, 1090 Vienna, Austria}%
%

\affiliation
{Faculty of Physics, University of Vienna, Boltzmanngasse 5, 1090 Vienna, Austria}%
%

\maketitle

\subsection{I. Introduction}

Since the seminal works of Einstein, Podolsky and Rosen (EPR)~\cite{Eins1935}
and Schrödinger~\cite{Schr1935} numerous experiments have demonstrated the
counter-intuitive effects of quantum entanglement, in particular the violation
of local realism through Bell's inequality~\cite{Bell1964}. Entanglement has
been demonstrated for many physical systems such as
photons~\cite{Free1972,Aspe1982,Kwia1995}, atoms~\cite{Hagl1997,Gros2011},
ions~\cite{Turc1998,Rowe2001}, and superconducting devices~\cite{Stef2006}.
Variants of the EPR experiment have been realized, for example exploiting the
analogy with quadrature phase operators \cite{Ou1992}. But, until now, nobody
was able to demonstrate the original EPR idea of an entangled state of freely
moving massive particles in their external degrees of freedom, i.e.\ a state
of the form $\int_{-\infty}^{\infty}$d$x\,\left\vert x\right\rangle
_{\text{A}}\left\vert x\!+\!x_{0}\right\rangle _{\text{B}}=\int_{-\infty
}^{\infty}$d$p\,$exp$(\tfrac{\text{i}}{\hbar}x_{0}p)\,\left\vert
p\right\rangle _{\text{A}}\left\vert -p\right\rangle _{\text{B}}$, where, $x$
and $p$ denote position and momentum, $x_{0}$ is a constant, and indices label
the two particles.

In this proposal, following the experimental approach of Ref.~\cite{Perr2007},
we consider a Bose--Einstein condensate (BEC) of metastable helium-4 ($^{4}%
$He$^{\ast}$). Via interactions with lasers, the particles are outcoupled from
the trap, brought to collision, and fall on a detector. The collisions prepare
atom pairs in a three-dimensional version of the EPR state with fixed absolute
momenta. While the idealized EPR state would have to be created everywhere in
space, our pairs are created in a finite space volume. We propose to test the
entanglement in a double double-slit experiment or via ghost interference.

We start with a short review of the procedure described in
Ref.~\cite{Perr2007}: $^{4}$He$^{\ast}$ atoms of mass $m\simeq6.646\!\cdot
\!10^{-27}\,$kg are magnetically trapped and form a cigar-shaped BEC along the
(horizontal) $x$ direction. One employs two $\sigma$-polarized laser beams
counterpropagating horizontally along the $+x$ and $-x$ directions and a $\pi
$-polarized laser beam from the top along the $-z$ direction fulfilling the
Raman condition, all with a wavelength $\lambda_{\text{L}}\simeq
1.083\,$\micro
m. They bring the atoms into a magnetically insensitive state and thereby
induce a velocity kick in the horizontal $\pm x$ directions and a kick upwards
along $+z$. The recoil velocity of each kick is $v_{\text{rec}}=\tfrac
{h}{\lambda_{\text{L}}\,m}\simeq92\,$mm$/$s, where $h$ is Planck's constant.
Thus, the three laser beams produce a superposition of two counterpropagating
matter waves of falling helium atoms, which subsequently scatter in a
four-wave mixing process. In total, a fraction of about 5\thinspace\% of all
$10^{4}$ to $10^{5}$ helium atoms collides and is scattered from the two
condensates. While the relative velocity of two scattered atoms is
$2\,v_{\text{rec}}$, the velocity uncertainties can be obtained from the
Gross-Pitaevskii equation. Assuming the trap parameters of
Ref.~\cite{Perr2007}, the BEC is elongated along $\pm x$ and the uncertainties
are anisotropic: $\Delta v_{x}\simeq0.0044\,v_{\text{rec}}$, $\Delta
v_{y,z}\simeq0.091\,v_{\text{rec}}$.

The collisions are, to a very good approximation, of s-wave type,
i.e.\ isotropic, and take place over a characteristic timescale of
$150\,$\micro s~\cite{Perr2007}. Depending on the size of the condensate the
collisions can produce momentum correlated particle pairs, lying on a shell in
velocity space, whose origin is at $v_{\text{rec}}\,\mathbf{\hat{e}}_{z}$ with
radius $v_{\text{rec}}$. Within quantum mechanical uncertainties, momentum
conservation requires the two partners to find themselves being anticorrelated
to each other in momentum space. Most importantly, the isotropic nature of the
s-wave scattering process gives rise to the superposition of all possible
emission directions and thus to quantum mechanical entanglement in the
external degrees of freedom of the two massive particles.

The atomic de Broglie wavelength $\lambda_{\text{dB}}$ associated with the
recoil velocity is the same as the wavelength of the laser beams:
$\lambda_{\text{dB}}=\tfrac{h}{m\,v_{\text{rec}}}\simeq1.083\,$\micro m. In
Refs.~\cite{Molm2008} and \cite{Perr2008} a model for the observed
Hanbury-Brown Twiss (collinear) and back-to-back (BB) correlations of
Ref.~\cite{Perr2007} is developed.

\subsection{II. Double double-slit experiment}

\begin{figure}[t]
\begin{center}
\includegraphics[width=7cm]{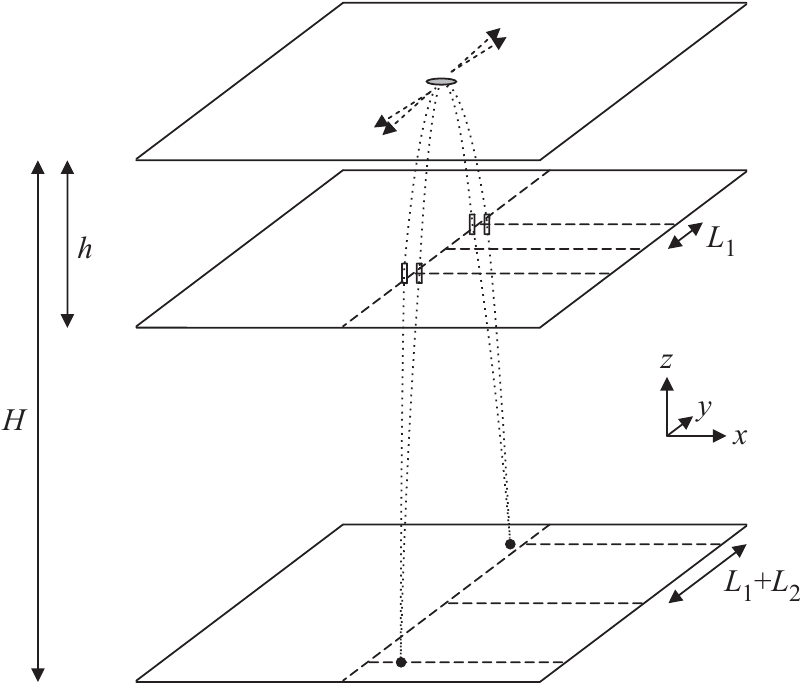}
\end{center}
\par
\vspace{-0.33cm}\caption{Schematic of the double double-slit experiment (not
drawn to scale). Pairs of atoms collide at height $H$ and fall under gravity
through the double slits onto the detector. See main text for details.}%
\label{fig setup}%
\end{figure}

Now consider a double double-slit arrangement as in Figure~\ref{fig setup}.
All particles hit by the lasers get a velocity kick of $v_{\text{rec}}$ in the
$\pm x$ direction as well as upwards along $+z$. Let us consider only those
particles which collided in such a way that they did not get any additional
vertical velocity component and are moving along $\pm y$ direction with
velocity $v_{\text{rec}}$ after the collision (i.e.\ along $\pm\mathbf{\hat
{e}}_{y}+\mathbf{\hat{e}}_{z}$ with velocity $\!\sqrt{2}\,v_{\text{rec}}$).
With gravity acceleration $g\simeq9.81\,$m$/$s$^{2}$ and distance to the
detector $H=0.5\,$m, the falling time is $\tau\simeq328.8\,$ms. Having the
coordinate origin $\mathbf{O}$ in the center of the initial condensate, the
atoms pass the double-slit at a lateral position of $y=L_{1}$ at some height
$-h$ and hit the detector at the maximally possible lateral distance
$y=L_{1}+L_{2}=v_{\text{rec}}\,\tau\simeq30.2\,$mm.

\begin{figure}[t]
\begin{center}
\includegraphics[width=.48\textwidth]{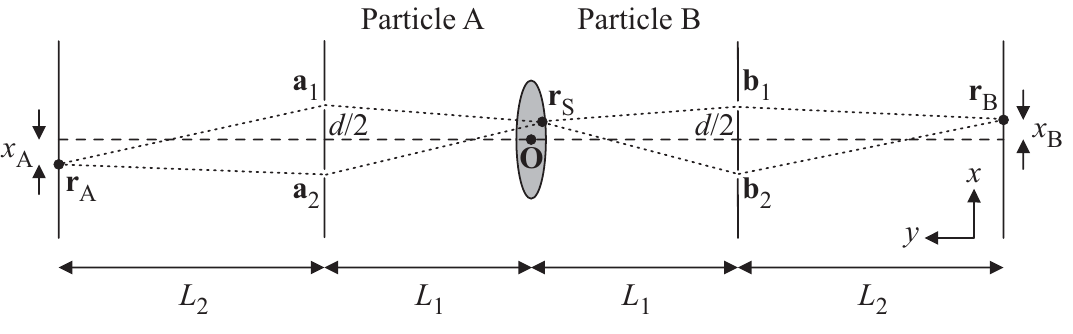}
\end{center}
\par
\vspace{-0.33cm}\caption{Schematic of the double double-slit experiment (top
view, not drawn to scale). Pairs of atoms, A and B, are emitted from source
points $\mathbf{r}_{\text{S}}$ within the BEC, pass double slits and arrive at
detectors at positions $\mathbf{r}_{\text{A}}$ and $\mathbf{r}_{\text{B}}$.
See main text for details.}%
\label{fig dds}%
\end{figure}

We can ignore the effect of gravity by considering only the top view of the
experiment (Figure~\ref{fig dds}). The source S of size $S_{x}\times
S_{y}\times S_{z}$ is emitting particle pairs A~and~B with
velocity~$v_{\text{rec}}$. At a distance $L_{1}$, there are two double-slits
with slit separation~$d$. At a further distance $L_{2}$ an observation screen
is located. There are four possible paths for getting a coincidence between
atom detections on the left and right side at positions $\mathbf{r}_{\text{A}%
}$ and $\mathbf{r}_{\text{B}}$ with $x$-coordinates $x_{\text{A}}$ and
$x_{\text{B}}$, respectively: \{$\mathbf{a}_{1}$,$\mathbf{b}_{1}$\},
\{$\mathbf{a}_{1}$,$\mathbf{b}_{2}$\}, \{$\mathbf{a}_{2}$,$\mathbf{b}_{1}$\},
and \{$\mathbf{a}_{2}$,$\mathbf{b}_{2}$\}. Here \{$\mathbf{a}_{i}$%
,$\mathbf{b}_{k}$\} means particle A passed slit $\mathbf{a}_{i}$ and particle
B passed slit $\mathbf{b}_{k}$ ($i,k=1,2$).

As discussed in Refs.~\cite{Horn1988,Gree1993,Horn1997}, there are two
limiting cases in a double double-slit experiment:

\begin{enumerate}
\item If the source size is very small, there is no momentum correlation
between the particles. The momentum spread of each individual particle is
large enough so that it can go through either slit at the same time and its
partner does not carry enough which-path-information to identify through which
of them it went. Thus, each particle forms a Young pattern independently of
its partner. One effectively has two single-particle interference patterns
unrelated to entanglement. The far field double double-slit (dds) two-particle
pattern at the observation screens is a product of two independent
one-particle patterns and is of the form%
\begin{equation}
|\psi_{\text{AB}}^{\text{(dds)}}(x_{\text{A}},x_{\text{B}})|^{2}\propto
\cos^{2}\!\left(  \pi\dfrac{x_{\text{A}}}{d_{f}}\right)  \cos^{2}\!\left(
\pi\dfrac{x_{\text{B}}}{d_{f}}\right)
\end{equation}
with $d_{f}$ being the fringe distance.

\item If the source size is large (but still small compared to the slit
distance), there is no one-particle interference pattern at either screen. The
large source implies a small two-particle momentum uncertainty and therefore a
high momentum correlation of the particle pairs. If one particle went through
one slit, the other particle must have gone through the diagonally opposite
one. However, since every particle is detected far behind a double-slit, the
which-path information about its partner is erased. There is a superposition
of two possibilities:\ particle A via upper slit \& particle B via lower slit,
and vice versa. The far field two-particle pattern cannot be factorized and
has the form%
\begin{equation}
|\psi_{\text{AB}}^{\text{(dds)}}(x_{\text{A}},x_{\text{B}})|^{2}\propto
\cos^{2}\!\left(  \pi\dfrac{x_{\text{A}}\!-\!x_{\text{B}}}{d_{f}}\right)  .
\end{equation}
The Young fringes on, say, side B can be seen only \textit{conditionally} on
finding particles at a certain detector position $x_{\text{A}}$ on the left
side. Only by measuring coincidences, an interference pattern can be seen. Its
maximum on the right side is at the same $x$-position as the detector at the
left side.
\end{enumerate}

Importantly, there is also a third regime, namely the one of very large source
size. If the condensate is comparable to or larger than the slit distance, the
two-particle interference pattern arising from the diagonal paths
\{$\mathbf{a}_{1}$,$\mathbf{b}_{2}$\} \& \{$\mathbf{a}_{2}$,$\mathbf{b}_{1}$\}
is superposed by a two-particle pattern originating from the horizontal paths
\{$\mathbf{a}_{1}$,$\mathbf{b}_{1}$\} \& \{$\mathbf{a}_{2}$,$\mathbf{b}_{2}%
$\}. Therefore, detection of one particle at either slit does not imply any
information about the slit the other particle goes through. This restores
one-particle interference and, due to complementarity, no genuine two-particle
interference arises~\cite{Jaeg1993}.

To calculate the two-particle interference pattern on the observation screen,
we follow the treatment in Ref.~\cite{Horn1997}, where one integrates over
point sources which emit two spherical waves without any (anti)correlation in
momentum. Remarkably, the anticorrelation and entanglement emerge naturally by
integrating spherical waves of two particles emitted from the same position
over a sufficiently large source area. This is in analogy to the Fourier
transformation of a single particle in phase space, giving rise to reduced
momentum uncertainty as the source grows larger. Let us denote the possible
path lengths from $\mathbf{r}_{\text{S}}\equiv(x_{\text{S}},y_{\text{S}%
},z_{\text{S}})$ to $\mathbf{r}_{\text{A}}\equiv(x_{\text{A}},L_{1}+L_{2},0)$
and $\mathbf{r}_{\text{B}}\equiv(x_{\text{B}},-L_{1}-L_{2},0)$ by
$L_{\mathbf{a}_{i}}\equiv\overline{\mathbf{r}_{\text{S}}\mathbf{a}_{i}%
}+\overline{\mathbf{a}_{i}\mathbf{r}_{\text{A}}}$ and $L_{\mathbf{b}_{i}%
}\equiv\overline{\mathbf{r}_{\text{S}}\mathbf{b}_{i}}+\overline{\mathbf{b}%
_{i}\mathbf{r}_{\text{B}}}$ with $i=1,2$, as calculated by simple geometry.
The (unnormalized) quantum mechanical amplitude for two entangled particles,
emerging from $\mathbf{r}_{\text{S}}$, to land at points $\mathbf{r}%
_{\text{A}}$ and $\mathbf{r}_{\text{B}}$ is%
\begin{align}
\psi_{\text{SAB}}^{\text{(dds)}}(\mathbf{r}_{\text{S}},\mathbf{r}_{\text{A}%
},\mathbf{r}_{\text{B}})  &  \propto\text{e}^{\text{i\thinspace}\frac{2\pi
}{\lambda_{\text{dB}}}\,(L_{\mathbf{a}_{1}}+L_{\mathbf{b}_{1}})}%
+\text{e}^{\text{i\thinspace}\frac{2\pi}{\lambda_{\text{dB}}}\,(L_{\mathbf{a}%
_{1}}+L_{\mathbf{b}_{2}})}\nonumber\\
&  \;\;\;\;+\text{e}^{\text{i\thinspace}\frac{2\pi}{\lambda_{\text{dB}}%
}\,(L_{\mathbf{a}_{2}}+L_{\mathbf{b}_{1}})}+\text{e}^{\text{i\thinspace}%
\frac{2\pi}{\lambda_{\text{dB}}}\,(L_{\mathbf{a}_{2}}+L_{\mathbf{b}_{2}})}.
\end{align}
It is a superposition of four equal-weight amplitudes corresponding to the
possible path combinations of particles A~and~B. We have omitted the
one-over-distance dependence of the amplitude of the spherical waves because
it is practically constant in the far field and can be taken into the
normalization factor. The phase of the amplitude of a collision to happen can
be treated constant over the source volume if one neglects the expansion of
the colliding BECs. Then, we can write down the quantum mechanical amplitude
for two entangled particles, emerging from the whole source, to land at points
$\mathbf{r}_{\text{A}}$~and~$\mathbf{r}_{\text{B}}$. It is a superposition of
all possible emission points over the source volume $V$ and is obtained via
integration over the whole condensate, possibly with some weighting function
$g(\mathbf{r}_{\text{S}})$:%
\begin{equation}
\psi_{\text{AB}}^{\text{(dds)}}(\mathbf{r}_{\text{A}},\mathbf{r}_{\text{B}%
})\propto\dfrac{1}{V}\,%
{\displaystyle\iiint\nolimits_{\text{S}}}
\,\text{d}\mathbf{r}_{\text{S}}\,g(\mathbf{r}_{\text{S}})\,\psi_{\text{SAB}%
}(\mathbf{r}_{\text{S}},\mathbf{r}_{\text{A}},\mathbf{r}_{\text{B}}).
\end{equation}

Summing up, we have the following conditions for a two-particle interference experiment:

\begin{enumerate}
\item[(I)] The source must be sufficiently large to achieve well defined
momentum correlation and wash out the single-particle interference pattern:%
\begin{equation}
\dfrac{\Delta p_{x}}{p}\ll\dfrac{d}{L_{1}}.
\end{equation}
The relative momentum spread $\tfrac{\Delta p_{x}}{p}=\tfrac{\Delta p_{x}%
}{mv_{\text{rec}}}$ must be small enough not to \textquotedblleft
illuminate\textquotedblright\ both slits. The source size $S_{x}$ implicitly
influences $\Delta p_{x}$. The larger the source size along $x$, the smaller
$\Delta p_{x}$ becomes and the better the condition can be fulfilled.

\item[(II)] The fringe distance $d_{f}$ should be much larger than the
detector resolution $\delta x$:%
\begin{equation}
d_{f}=\lambda_{\text{dB}}\,\dfrac{L_{2}}{d}>5\,\delta x.
\end{equation}
Here we assume that one needs at least 5 pixels per oscillation.

\item[(III)] The source must be sufficiently small not to destroy the
two-particle interference pattern:%
\begin{equation}
S_{x}\ll d. \label{eq cond III}%
\end{equation}
If the source size becomes comparable to or larger than the slit distance, two
two-particle interference patterns wash each other out.
\end{enumerate}

Now we come to the parameter analysis. We take $\tfrac{\Delta p_{x}}{p}%
=\frac{1}{\pi}\sqrt{\frac{21}{8}}\,\frac{\lambda_{\text{dB}}}{S_{x}}%
\simeq\frac{0.56\,\micro\text{m}}{S_{x}}$ \cite{Sten1999} and $S_{y}%
=10\,\micro$m. This leaves $S_{x}$ as a free parameter, and we can write
conditions (I), (II), and (III) in a single line:%
\begin{equation}
0.56\,\micro\text{m}\times\frac{L_{1}}{d}\;\overset{(\text{I})}{\ll}%
\;S_{x}\;\overset{(\text{III})}{\ll}\;d\;\overset{(\text{II})}{<}%
\;\frac{\lambda_{\text{dB}}\,L_{2}}{5\,\delta x}.\label{eq conditions}%
\end{equation}
An approximate solution for all our simultaneous constraints is
$d=100\,\micro$m, $L_{1}=5\,$mm, $L_{2}=25\,$mm. Here $L_{1}+L_{2}$ exploits
the maximal possible lateral distance given by the distance between BEC and
detector of $H=0.5\,$m. The fringe distance becomes $d_{f}\simeq271\,\micro$m,
which means that a fringe is resolved by only 4 to 5 pixels, given a detector
resolution of $\delta x\simeq60\,$\micro m.\begin{figure}[t]
\begin{center}
\includegraphics[width=.48\textwidth]{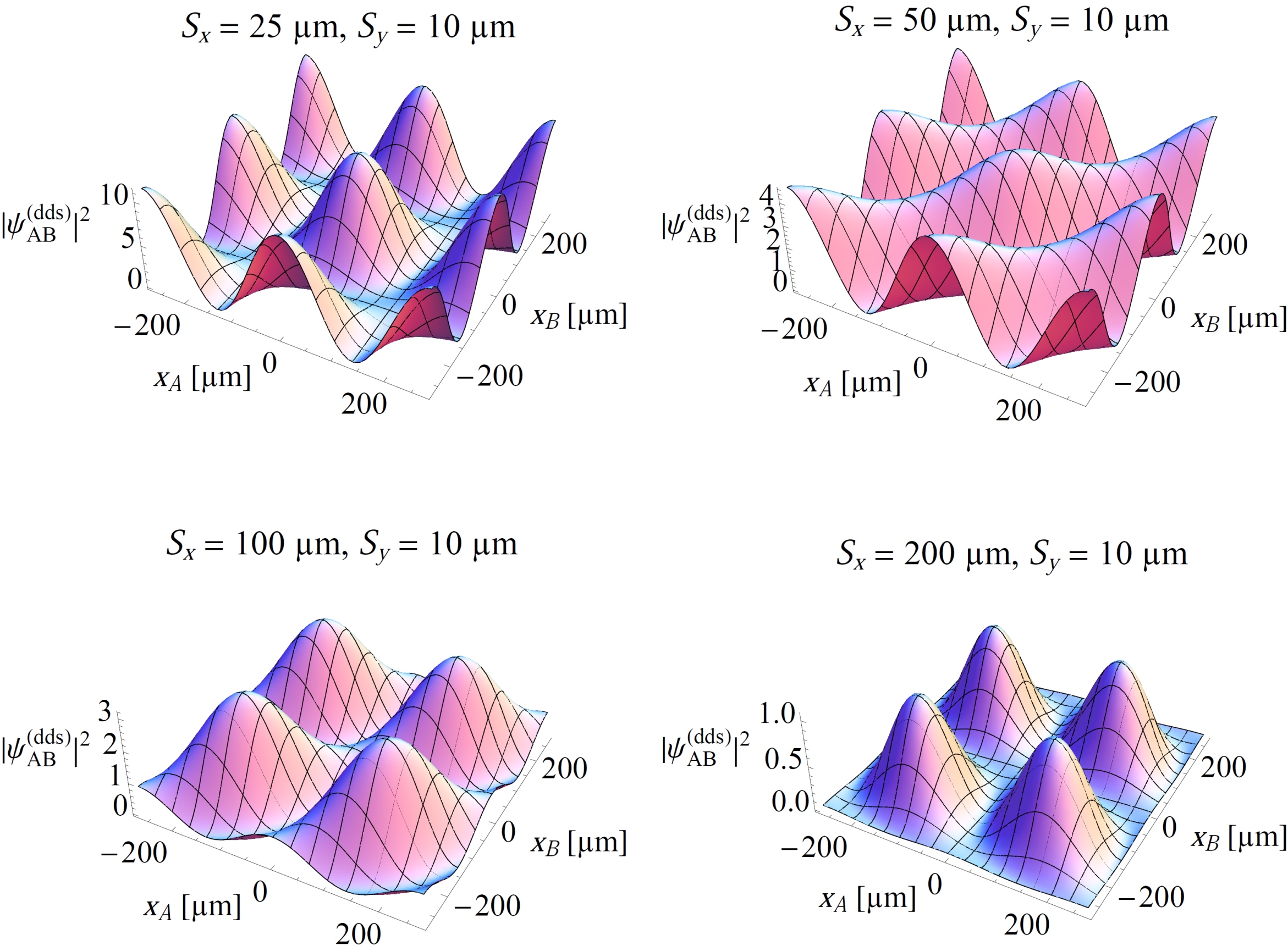}
\end{center}
\par
\vspace{-0.33cm}\caption{(Color online.) Two-particle probability distribution
$|\psi_{\text{AB}}^{\text{(dds)}}|^{2}$ for the double double-slit experiment
for different source sizes. The slit distance $d=100\,\micro$m, source-slit
distance $L_{1}=5\,$mm, and slit-detector distance $L_{2}=25\,$mm, are kept
fixed, resulting in a constant fringe distance $d_{f}\simeq271\,\micro$m. For
a very small source (top left, $S_{x}=25\,\micro$m) the momentum spread of
each individual particle is large and one obtains a product of two
one-particle patterns of the form $\cos^{2}(\pi\tfrac{x_{\text{A}}}{d_{f}%
})\cos^{2}(\pi\tfrac{x_{\text{B}}}{d_{f}})$. If the source is larger than the
slit distance (bottom right, $S_{x}=200\,\micro$m), two two-particle patterns
together wash out again into a factorizable pattern. An intermediate source
size (top right, $S_{x}=50\,\micro$m) fulfills all conditions for two-particle
interference and shows a distribution of the unfactorizable form $\cos^{2}%
(\pi\tfrac{x_{\text{A}}-x_{\text{B}}}{d_{f}})$.}%
\label{fig optimization}%
\end{figure}

Figure \ref{fig optimization} shows $|\psi_{\text{AB}}^{\text{(dds)}}|^{2}$
for various source sizes $S_{x}$. (The integration was done over a
two-dimensional rectangular source with size $S_{y}=10\,$\micro m along $y$
and constant weighting function. Integration over $y$ only marginally changes
the pattern and so would an integration along $z$.) While in the top left
picture ($S_{x}=25\,\micro$m) condition (I) is violated, the bottom pictures
($S_{x}=100\,\micro$m and $S_{x}=200\,\micro$m) violate condition (III). The
top right case ($S_{x}=50\,$\micro m) shows a conditional interference pattern
with high genuine two-particle visibility. For this choice of parameters
conditions (\ref{eq conditions}) read $28\,\micro$m$\;\ll\;50\,\micro$%
m$\;\ll\;100\,\micro$m$\;<\;90\,\micro$m and are \textit{approximately}
fulfilled. Decreasing or increasing the source size just a bit, lets us run
into one or the other limitation. This shows that there is essentially no
further freedom in any of the parameters given typical experimental
constraints (falling height $H$, detector resolution $\delta x$). A double
double-slit experiment therefore requires careful adjustment of the
experimental parameters.

\subsection{III. Ghost interference}

\begin{figure}[t]
\begin{center}
\includegraphics[width=.48\textwidth]{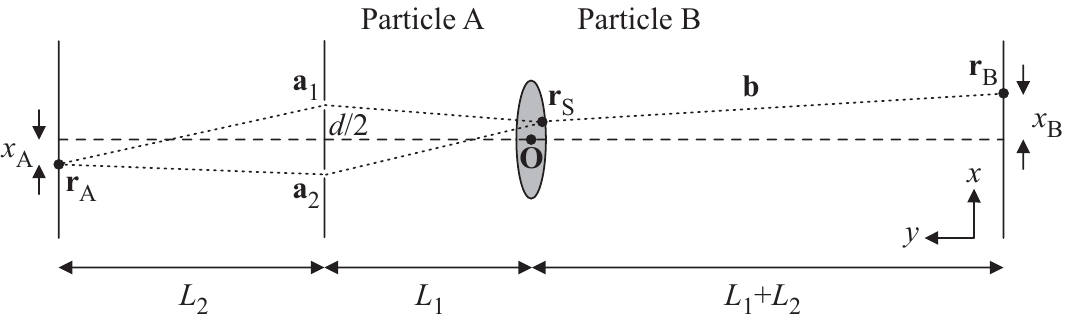}
\end{center}
\par
\vspace{-0.33cm}\caption{Schematic of the ghost interference experiment (top
view, not drawn to scale). Pairs of atoms, A and B, are emitted from source
points~$\mathbf{r}_{\text{S}}$ within the BEC and arrive at detectors at
positions $\mathbf{r}_{\text{A}}$ and $\mathbf{r}_{\text{B}}$. Only atom A
passes through a double slit. See main text for details.}%
\label{fig ghost}%
\end{figure}

However, it turns out that we can circumvent condition (III),
Ineq.\ (\ref{eq cond III}), by removing one of the double-slits and using a
ghost interference setup \cite{Stre1995}, as shown in Figure \ref{fig ghost}%
.

The possible path lengths from $\mathbf{r}_{\text{S}}$ to $\mathbf{r}%
_{\text{A}}\equiv(x_{\text{A}},L_{1}+L_{2},0)$ and $\mathbf{r}_{\text{B}%
}\equiv(x_{\text{B}},-L_{1}-L_{2},0)$ are abbreviated as $L_{\mathbf{a}_{i}%
}\equiv\overline{\mathbf{r}_{\text{S}}\mathbf{a}_{i}}+\overline{\mathbf{a}%
_{i}\mathbf{r}_{\text{A}}}$ ($i=1,2$) and $L_{\mathbf{b}}\equiv\overline
{\mathbf{r}_{\text{S}}\mathbf{r}_{\text{B}}}$. The quantum mechanical
amplitude for two entangled particles in ghost interference (gh), emerging
from point $\mathbf{r}_{\text{S}}$, to land at points $\mathbf{r}_{\text{A}}$
and $\mathbf{r}_{\text{B}}$ is%
\begin{equation}
\psi_{\text{SAB}}^{\text{(gh)}}(\mathbf{r}_{\text{S}},\mathbf{r}_{\text{A}%
},\mathbf{r}_{\text{B}})\propto\text{e}^{\text{i\thinspace}\frac{2\pi}%
{\lambda_{\text{dB}}}\,(L_{\mathbf{a}_{1}}+L_{\mathbf{b}})}+\text{e}%
^{\text{i\thinspace}\frac{2\pi}{\lambda_{\text{dB}}}\,(L_{\mathbf{a}_{2}%
}+L_{\mathbf{b}})}.
\end{equation}
The quantum mechanical amplitude for two entangled particles, emerging from
the whole source, to land at points $\mathbf{r}_{\text{A}}$~and~$\mathbf{r}%
_{\text{B}}$ is again given by integration over all point sources:%
\begin{equation}
\psi_{\text{AB}}^{\text{(gh)}}(\mathbf{r}_{\text{A}},\mathbf{r}_{\text{B}%
})\propto\dfrac{1}{V}\,%
{\displaystyle\iiint\nolimits_{\text{S}}}
\,\text{d}\mathbf{r}_{\text{S}}\,g(\mathbf{r}_{\text{S}})\,\psi_{\text{SAB}%
}^{\text{(gh)}}(\mathbf{r}_{\text{S}},\mathbf{r}_{\text{A}},\mathbf{r}%
_{\text{B}}).
\end{equation}
In the ghost interference setup with only one double-slit, condition (III) is
not necessary any longer. The remaining conditions (I) and (II) in
(\ref{eq conditions}) can be easily fulfilled. Let us for instance choose
$d=50\,\micro$m, $L_{1}=5\,$mm, $L_{2}=25\,$mm. Due to the smaller slit
distance as compared to the double double-slit scenario, the fringe distance
on side A becomes $d_{f}^{\text{(A)}}=\lambda_{\text{dB}}\,\tfrac{L_{2}}%
{d}\simeq542\,\micro$m. The fringe distance on side B can be calculated via
elementary geometrical considerations. It is $d_{f}^{\text{(B)}}%
=\lambda_{\text{dB}}\,\tfrac{2\,L_{1}+L_{2}}{d}\simeq758\,\micro$m. Both can
easily be resolved with modern detectors. Conditions (I) and (II) read
$S_{x}\gg56\,\micro$m, $d=50\,\micro$m$\;\ll83\,\micro$m, leaving the only
constraint $S_{x}\gg56\,$\micro m.

\begin{figure}[t]
\begin{center}
\includegraphics[width=.48\textwidth]{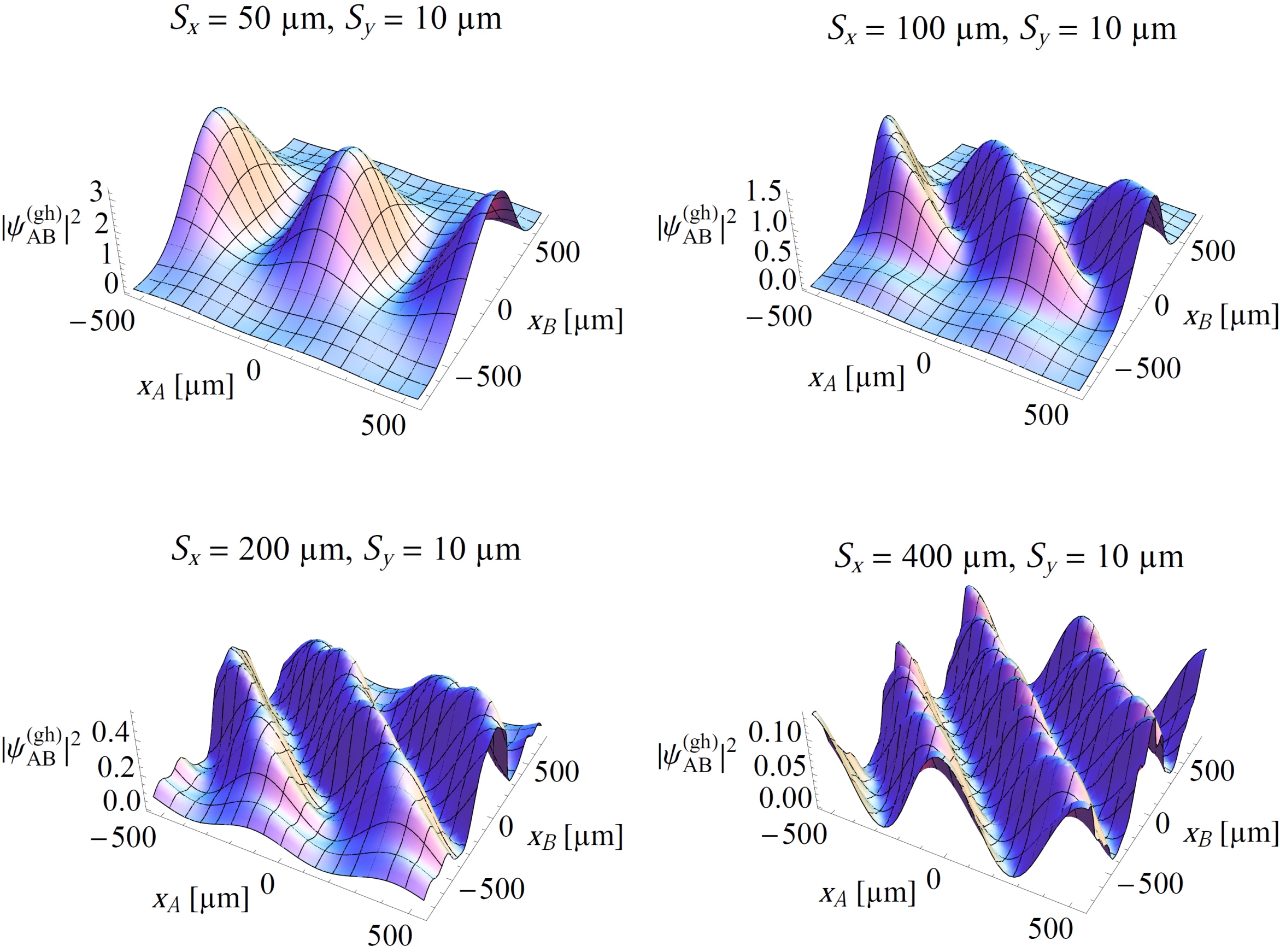}
\end{center}
\par
\vspace{-0.33cm}\caption{(Color online.) Two-particle probability distribution
$|\psi_{\text{AB}}^{\text{(gh)}}|^{2}$ for the ghost interference experiment
for different source sizes. The source-slit distance $L_{1}=5\,$mm and
slit-detector distance $L_{2}=25\,$mm are the same as in
Figure~\ref{fig optimization}. The slit distance $d=50\,\micro$m is smaller
now, leading to larger fringe distances $d_{f}^{\text{(A)}}\simeq542\,\micro$m
and $d_{f}^{\text{(B)}}\simeq758\,\micro$m at side A and B, respectively. A
very small source (top left, $S_{x}=50\,\micro$m) results in a product of two
one-particle patterns. Large sources (bottom, $S_{x}=200\,\micro$m and
$400\,\micro$m) show two-particle interference and produce a pattern of the
unfactorizable form $\cos^{2}(\pi\tfrac{x_{\text{A}}+x_{\text{B}}}{d_{f}})$.}%
\label{fig ghost_comparison}%
\end{figure}

Figure \ref{fig ghost_comparison} shows $|\psi_{\text{AB}}^{\text{(gh)}}|^{2}$
for different source sizes. One can see the transition from one-particle
interference ($S_{x}=50\,$\micro m, condition (I) not fulfilled) to almost
perfect two-particle interference ($S_{x}=400\,$\micro m) of the form
$\cos^{2}(\pi\tfrac{x_{\text{A}}+x_{\text{B}}}{d_{f}})$. Note that condition
(III), $S_{x}\ll d$, is clearly violated for the larger source sizes.
Moreover, in contrast to the double double-slit experiment, the maximum of the
interference pattern on one side is opposite to the conditioning detection on
the other side.

Recently it was reported in Ref.~\cite{Kher2012} that matter waves from
colliding BECs violate the Cauchy-Schwarz inequality, ruling out a description
in terms of classical stochastic random variable theories. The two-particle
patterns discussed in the present work cannot be explained by classical
correlations as soon as the two-particle visibility exceeds $\frac{1}{2}$
\cite{Ghos1987,Beli1992}. A strict proof that they must arise due to quantum
entanglement can be made by employing a separability criterion using modular
variables~\cite{Gnei2011}.

\subsection{IV. Pair identification}

Finally, we come to an important issue, namely the problem of identifying
coincidences. The pulse duration of the lasers is of the order of
500\thinspace ns, which is negligible. The traveling time of the recoiled
atoms from one side of the condensate to the center, i.e.\ a distance of about
$100\,\micro$m (more, if the condensate size is increased), is 1\thinspace ms.
Therefore, all collisions certainly happen within $\Delta\tau_{\text{coll}%
}\simeq1\,$ms, most of them likely within a fraction of that time.
Ref.~\cite{Perr2007} states that the time constant in the decay of the
collision rate is $150\,$\micro s. As discussed above, let us consider a pair
of particles which collide into the $+y$ and $-y$ direction, with unchanged
velocity component $v_{\text{rec}}$ along $+z$. Due to the velocity
uncertainty we should assume that one particle has a velocity along $z$ of
$v_{\text{rec}}$ and its entangled partner has $v_{\text{rec}}+\Delta v_{z}$.
The times when they hit the detector plate are $\tau_{v_{\text{rec}}}%
\simeq328.8\,$ms and $\tau_{v_{\text{rec}}+\Delta v_{z}}\simeq329.7\,$ms. This
means that two particles which must experimentally be identified as a
coincidence can have a time spread of $\Delta\tau_{\text{pair}}\simeq0.9\,$ms,
which is of the order of the collision time scale $\Delta\tau_{\text{coll}}$.
Two-particle interference requires to measure coincidences of entangled
partners, i.e.\ identification of the correct subensemble on side A
conditional on detection on side B. Since an identification by arrival time
seems to be impossible, it appears to be necessary to reduce the laser
intensity such that per shot, on average, only a few pairs collide in a way
that they can reach the maximal lateral distance $L_{1}+L_{2}$ on opposite
sides of the detector. Taking into account the finite detection efficiency
makes the pair identification experimentally extremely challenging. It may be
advantageous to replace the double slit (in ghost interference) by a grating
to increase the possible count rates.

\subsection{V. Conclusion}

An experimental demonstration of the original EPR gedanken experiment using
momentum entanglement between pairs of two colliding BECs is within reach.
While the problem of pair identification is very challenging, the constraints
of source size, time-of-flight distances, and detector resolution are
manageable and let conditional two-particle interference in a double
double-slit configuration or a ghost interference setup seem feasible.

\subsection{Acknowledgments}

The research was funded by the Doctoral Program CoQuS (Grant W1210) and
SFB-FoQuS of the Austrian Science Fund (FWF).

\end{document}